\documentclass[aps,pra,tightenlines,twocolumn,superscriptaddress,floatfix]{revtex4}

\usepackage{amssymb}
\usepackage{graphicx}
\usepackage{amsmath}
\usepackage{amsbsy}
\usepackage{amsthm}
\usepackage{bbm}
\usepackage{epsfig}

\newcommand{\beq}{\begin{eqnarray}}
\newcommand{\eeq}{\end{eqnarray}}

\newcommand{\ket}[1]{\ensuremath{| #1 \rangle}}
\newcommand{\bra}[1]{\ensuremath{\langle #1 |}}

\begin{document}

\title{Genuine Multipartite Entanglement in the $3$-Photon Decay of Positronium}
\author{Beatrix C. Hiesmayr}
\affiliation{Faculty of Physics, University of Vienna, Boltzmanngasse 5, 1090 Vienna, Austria}
\author{Pawel Moskal}
\affiliation{Institute of Physics, Jagiellonian University, Cracow, Poland}

\begin{abstract}
The electron-positron annihilation into two photons is a standard technology in medicine to observe e.g. metabolic processes in human bodies. A new tomograph  will provide the possibility to observe not only direct $e^+ e^-$ annihilations but also the $3$ photons from the decay of ortho-positronium atoms formed in the body. We show in this contribution that the three-photon state with respect to polarisation degrees of freedom depends on the angles between the photons and exhibits various specific entanglement features. In particular genuine multipartite entanglement, a type of entanglement involving all degrees of freedoms, is subsistent if the positronium was in a definite spin eigenstate. Remarkably, when all spin eigenstates are mixed equally, entanglement --and even stronger genuine multipartite entanglement-- survives. Due to a ``\textit{symmetrization}'' process, however, $Dicke$-type of entanglement remains whereas $GHZ$-type of entanglement vanishes. The survival of particular entanglement properties in the mixing scenario may make it possible to extract quantum information in form of distinct entanglement features, e.g., from metabolic processes in human bodies.
\end{abstract}

\maketitle


\section{Introduction}

The detection of the two high energetic photons coming from the annihilation of an electron and a positron is a well-established successful technology to image metabolic processes in living bodies (PET: Positron Emission Tomography). PET application are used in many different fields of medicine, e.g. in oncology, in cardiology, in radiation therapy or in neurology. In recent years, PET instrumentation has undergone a steady multifaceted evolution and the improvements include new hardware, new reconstruction methods and implementation of time-of-flight techniques~\cite{PETdevelopment,PETdevelopment2,PETdevelopment3,PETdevelopment4,PETdevelopment5,PETdevelopment6,PETdevelopment7}.

With no doubt PET serves as an important tool in imaging metabolic processes based on the sensitivity to tracers (positron-emitting radionuclides) injected into the body or tissue.

\begin{figure}[t]
\centering{\includegraphics[width=0.33\textwidth]{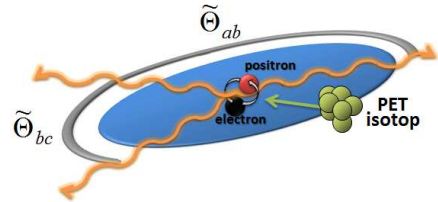}}
\caption{(Color online) This graphic shows schematically how from an isotop typically used in standard PET-therapy, e.g. FDG-$18$ (fludeoxyglucose), positronium is generated that decays into three photons which wave vectors have to lay in one plane due to energy and momentum conservation.}
	\label{graphicphotons}
\end{figure}

Electron-positron annihilations may occur either directly or via the creation of positronium atoms (a bound state of electron and positron). Positronium~\cite{positronium1,positronium2,positronium3,positronium4} can be in an anti-symmetric spin state (para-positronium) or a symmetric spin state (ortho-positronium). Charge conjugation implies that in the first case it decays into an even number of photons ($2\gamma$, $4\gamma$,\dots) and in the other case into an odd number of photons ($3\gamma$, $5\gamma$,\dots).  Due to kinematics and smallness of the fine-structure constant the $2\gamma$ and $3\gamma$ cases are the two most likely options. Since positronium atoms are formed copiously inside the human body during routine PET imaging $3\gamma$-decays occur also frequently. Even in water the production of ortho-positronium has a probability equal to about $25$\%~\cite{PET3gammeprobability} and is expected to be more than 38\% in a tissue~\cite{tissueprobability}. Three-photon events, however, have never been used in tomography because of technical limitations of standard PET devices. A new prototype, called J-PET (Jagiellonian-PET)~\cite{JPET1,JPET2,JPET3,PawelPatent,Dariajpet,JPETGajus}, has shown to meet all technical requirements in performing such a measurement for the first time.

This paper investigates the entanglement in the polarisation degrees of freedom of the three photons resulting from the decay of the ortho-positronium. Both for a fixed spin quantization direction of the positronium as well as the case of equal mixing. Photons are fascinating quantum systems, having spin one, but due to their mass-less property there is a nontrivial coupling between the spin and momentum properties. The most appropriate single-photon description remains controversial. A recent framework describing all single-photon states and single-photon observables by POVMs (positive-operator valued measurements) can be found in Ref.~\cite{Giacomo}. In this contribution we restrict ourselves to the polarisation degrees of freedom and  are interested in the correlation of three photons with energies that ranges from $0$ to $511$keV. Entanglement and in particular multipartite entanglement is a highly investigated field that has the potential to become a new technology. This paper makes a step towards investigating what type of entanglement is present in the three-photon state generated by the decay of ortho-positronium. This may one day result in obtaining not only the local information where in a tissue the positronium decays, but as well revealing the quantum information which may serve as a new quantum marker for specific biological processes.

Note that entanglement seems to play an important role in biological systems as e.g. observed in the light harvesting complexes, e.g. Ref.~\cite{bio1}, in bird navigation (European robin)~\cite{bio2,bio3} or in olfaction~\cite{bio4}. Let us emphasize here that these works have led to a paradigms change concerning that life may be too ``warm and wet'' for quantum phenomena to endure.

The paper is organized as follows. Section~\ref{state} introduces the $3$-photon state resulting from the decay of ortho-positronium. The next Section~\ref{pure} analyses the multipartite entanglement of the pure state scenario, followed by Section~\ref{reduced} discussing the distribution of entanglement among the three photons. The next Section~\ref{mixedness} shows that entanglement is not lost even in the mixed scenario. This is followed by a summary and outlook.

\section{The states relevant in ortho-positronium decays}\label{state}

As shown in Ref.~\cite{JPETEntanglementGeneration} the $3$-photon state from ortho-positronium decays for a fixed quantization direction $\hat{\vec{n}}$ of the positronium having a zero spin third component $s_{\hat{\vec{n}}}=0$ derives to
\begin{widetext}
\beq
|\Psi_{s_{\hat{\vec{n}}}=0}\rangle &=&\frac{1}{\sqrt{N}} \left(\cos(\Phi_{plane})\mathbbm{1}^{\otimes 3}+\sin(\Phi_{plane})\sigma_x^{\otimes 3}\right)\cdot\hat{\mathcal{R}}_{pol}(\tilde\Theta_{ab},\tilde\Theta_{bc}) \;|\Psi\rangle_{abc}
\eeq
with the normalisation
\beq N=\frac{1}{2} \left(9+\cos2\tilde\Theta_{ab}+\cos(2\tilde\Theta_{ab}+2\tilde\Theta_{bc})+\cos2\tilde\Theta_{bc}-4(\cos\tilde\Theta_{ab}+\cos(\tilde\Theta_{ab}+\tilde\Theta_{bc})
+\cos\tilde\Theta_{bc})\right)\;.
\eeq
\end{widetext}
Here the angle $\Phi_{plane}\in[0,\frac{\pi}{2}]$ is the angle between the quantization direction $\hat{\vec{n}}$ of the positronium and the decay plane formed by the momentum vectors of the three photons (momentum conservation forces the three momenta to lay in one plane), see Fig.~\ref{graphicphotons}. The operator $\hat{\mathcal{R}}_{pol}(\tilde\Theta_{ab},\tilde\Theta_{bc})$ covers the symmetries superposed by the decay process on the polarisation, where $\tilde\Theta_{ij}$ corresponds to the angles between photon $i$ and $j$, that all lay in the decay plane. The restriction due to momentum and energy conservation onto these angles is discussed at the end of this section. The operator is defined by
\begin{widetext}
\beq
\hat{\mathcal{R}}_{pol}(\tilde\Theta_{ab},\tilde\Theta_{bc})=\sum_{i,j,k=0}^{1}\left( (-1)^k \sin^2\frac{\tilde\Theta_{ab}}{2}+(-1)^j \sin^2(\frac{\tilde\Theta_{ab}}{2}+\frac{\tilde\Theta_{bc}}{2})+(-1)^i \sin^2\frac{\tilde\Theta_{bc}}{2}\right)|ijk\rangle_{abc}\langle ijk|_{abc}\;.
\eeq
\end{widetext}
This operator is invariant under permutation of the three photons such as the state (without normalisation)
\beq\label{reinerzustand}
|\Psi\rangle_{abc}&=&|000\rangle_{abc}-|110\rangle_{abc}-|011\rangle_{abc}-|101\rangle_{abc}\nonumber\\
&=&|\phi^-\rangle_{ab}\otimes |0\rangle_c-|\psi^+\rangle_{ab}\otimes |1\rangle_c\nonumber\\
&=& |0\rangle_a\otimes |\phi^-\rangle_{bc}- |1\rangle_a\otimes|\psi^+\rangle_{bc}\nonumber\\
&=&|\phi^-\rangle_{ac}\otimes |0\rangle_b-|\psi^+\rangle_{ac}\otimes |1\rangle_b\nonumber\\
&=&|+++\rangle_{abc}+|---\rangle_{abc}
\eeq
written in the computational basis $\{|0\rangle,|1\rangle\}$ which are defined as the eigenstates with respect to internal frame of each photon and may be identified with the linear polarised states $|H\rangle,|V\rangle$. The states $|\phi^\pm\rangle=|00\rangle\pm|11\rangle, |\psi^\pm\rangle=|01\rangle\pm|10\rangle$ are the Bell states (not normalized).
The states $|+/-\rangle$ correspond to the right/left handed circular polarised photons with respect to the choice of internal space of each photon $|\pm\rangle\;=\;\frac{1}{\sqrt{2}}\{|0\rangle\pm i |1\rangle\}$.
Assuming that a particular photon $i$ travels in $z$-direction, then $|0\rangle,|1\rangle$ can be identified also with the electric field components in $x,y$-direction, respectively.

\begin{widetext}
The two other possible eigenstates of the ortho-positronium having total spin one are $s_{\hat{\vec{n}}}=\pm 1$ are obtained by three local rotations, i.e.
\beq
|\Psi_{s_{\hat{\vec{n}}}=+1}\rangle &=&\sigma_x^{\otimes 3}\; |\Psi_{s_{\hat{\vec{n}}}=0}\rangle\nonumber\\
&=&\frac{1}{\sqrt{N}} \left(\sin(\Phi_{plane})\mathbbm{1}^{\otimes 3}+\cos(\Phi_{plane})\sigma_x^{\otimes 3}\right)\cdot\hat{\mathcal{R}}_{pol}(\tilde\Theta_{ab},\tilde\Theta_{bc}) \;|\Psi\rangle_{abc}\;,\nonumber\\
|\Psi_{s_{\hat{\vec{n}}}=-1}\rangle &=&\sigma_y^{\otimes 3}\; |\Psi_{s_{\hat{\vec{n}}}=0}\rangle\nonumber\\
&=&\frac{1}{\sqrt{N}} \left(\sin(\Phi_{plane})\sigma_z^{\otimes 3}+\cos(\Phi_{plane})\sigma_y^{\otimes 3}\right)\cdot\hat{\mathcal{R}}_{pol}(\tilde\Theta_{ab},\tilde\Theta_{bc}) \;|\Psi\rangle_{abc}\;.
\eeq
These states equal in the case of $\Phi_{plane}=0$ up to overall phases the result presented in Ref.~\cite{Acin}.
\end{widetext}

\begin{figure}
\centering{\includegraphics[width=0.3\textwidth]{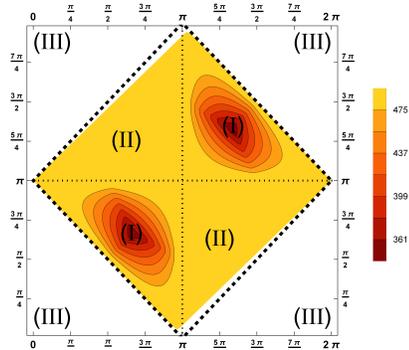}
}
\caption{(Color online) These contour plots show the maximum taken over all single photon energies of three photons for the allowed angles $\tilde\Theta_{ab}$ ($x$-axis) and $\tilde\Theta_{bc}$ ($y$-axis). Three kinematically different regions emerge. A forbidden region (III) where the momentum conservation does not hold since all wave vectors point into one half of the plane. Another region (II) where two photons have the maximum or minimum possible energy $\frac{E}{2}$, another physically forbidden region. And the region (I) where the energies are not extremal. This plot agrees with the Fig.~8 of Ref.~\cite{Dariajpet}, where also a Dalitz plot is shown for this case.}
	\label{allowedangles2}
\end{figure}

Due to momentum and energy conservation we have additional constraints regarding the two angles $\tilde\Theta_{ab},\tilde\Theta_{bc}$. Energy conservation in the rest mass system of the positronium leads to ($\hbar c\equiv 1$)
\beq
\omega_a+\omega_b+\omega_c=E\;,
\eeq
which fixes one energy of the three photons. Momentum conservation relates the energies of two photons to the solid angle between the three momenta, i.e.
\beq
\cos\tilde\Theta_{ab}=\frac{\frac{1}{2}-\frac{\omega_a}{E}-\frac{\omega_b}{E}+\frac{\omega_a\omega_b}{E^2}}{\frac{\omega_a \omega_b}{E^2}}\;.
\eeq
A solution is only obtained if right hand side is in the interval $[-1,1]$. The lower bound $-1$ implies that a single photon can have at most half of the total energy $E$ and the upper bound $+1$ bounds the sum of both energies to half of the total energy $E$. The possible range of angles are shown in Fig.~\ref{allowedangles2}. The kinematics thus singles out the region denoted by $(I)$, i.e. not the full parameter space is physically attainable due to energy and momentum conservation.

\section{Entanglement properties of the pure $3$-photon-states}\label{pure}

\begin{figure*}
\centering{(a)\includegraphics[width=0.3\textwidth]{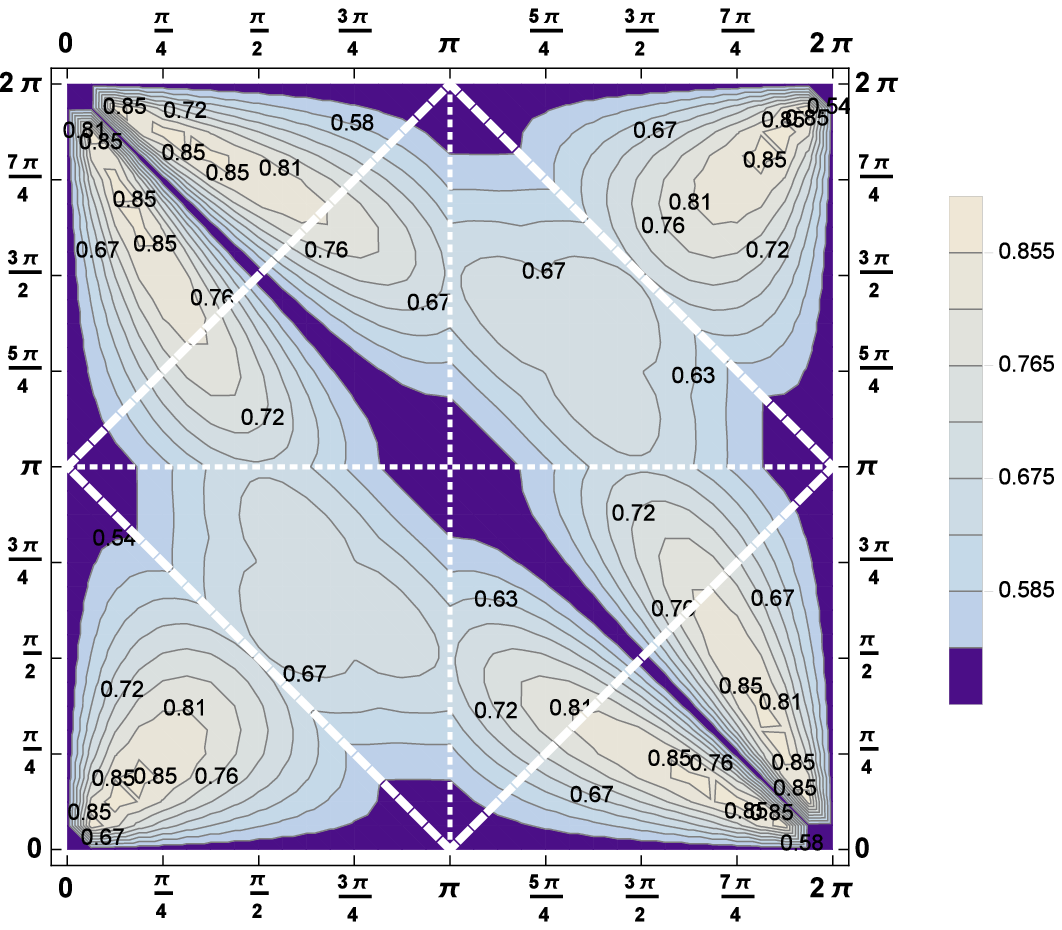}
(b)\includegraphics[width=0.3\textwidth]{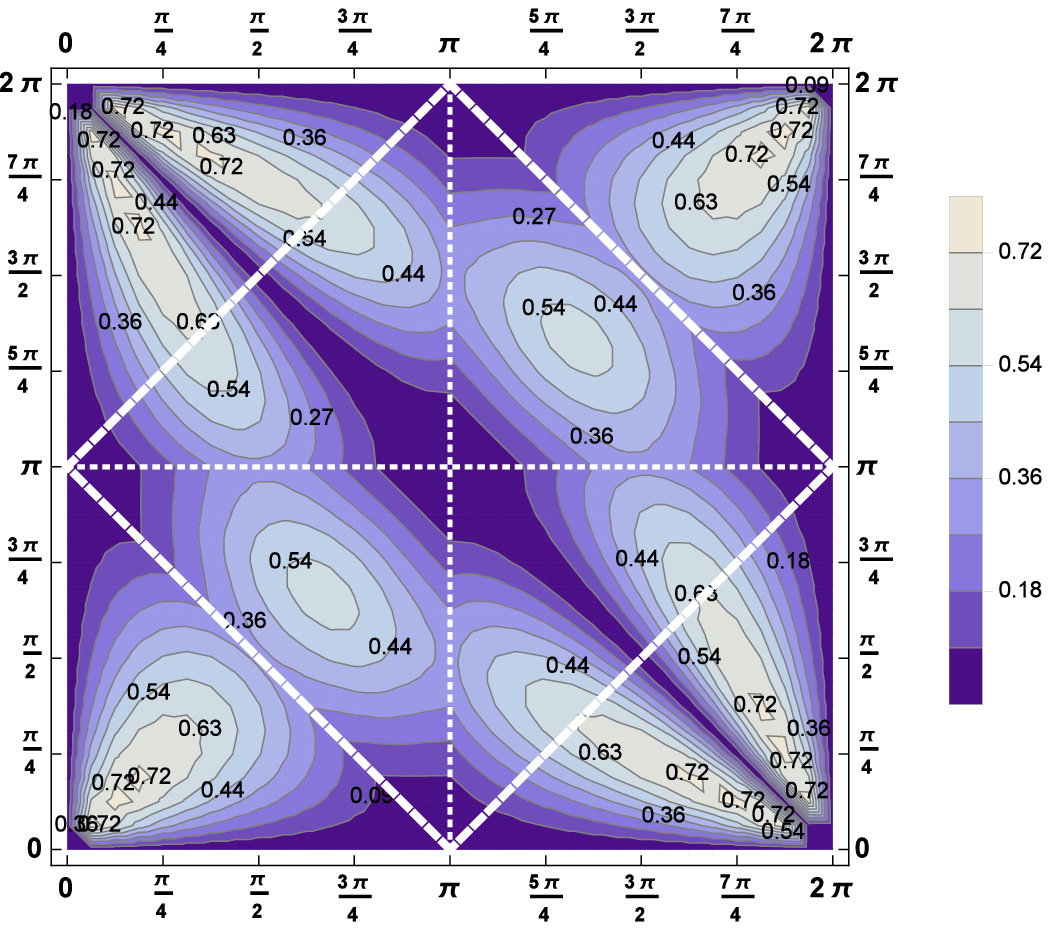}
(c)\includegraphics[width=0.3\textwidth]{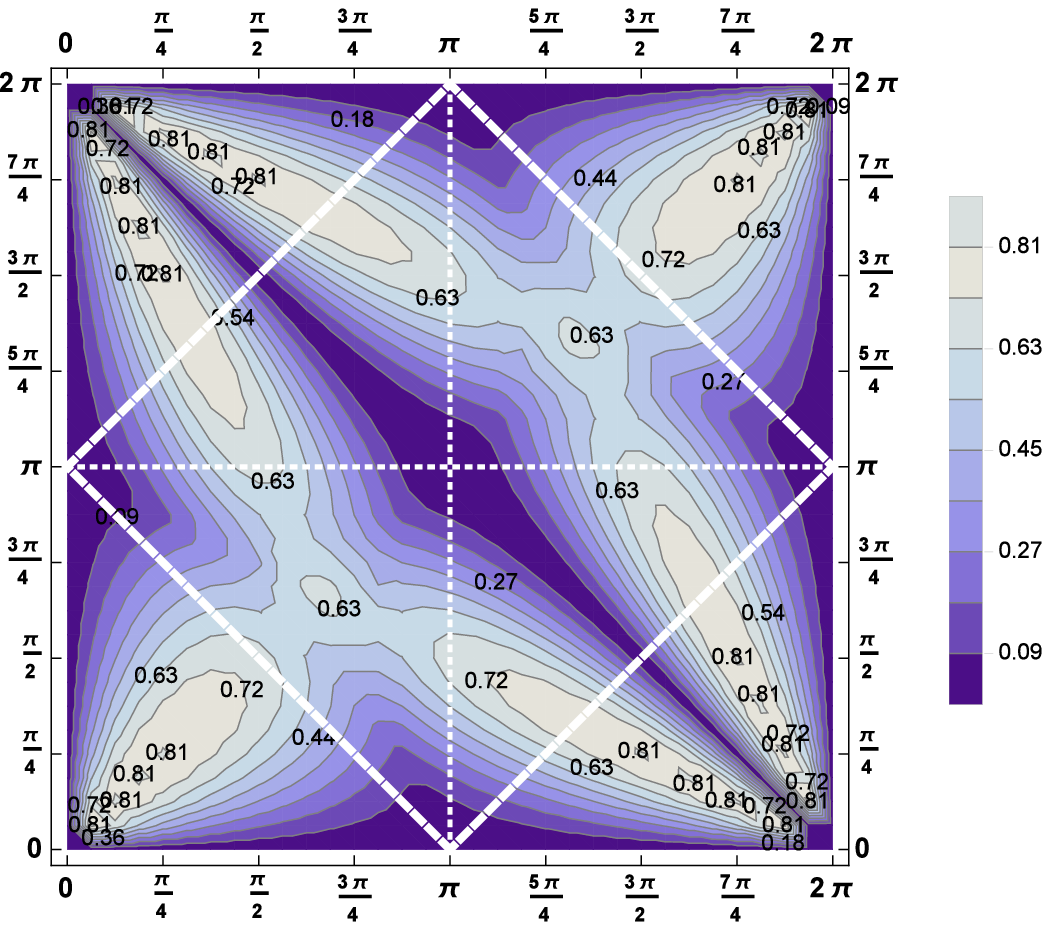}}
\caption{(Color online) These contour plots show the function (a) $Q_{SEP}$, (b) $Q_{GHZ}$ and (c) $Q_{W}$ for the pure state $|\psi_{pure}(\tilde\Theta_{ab},\tilde\Theta_{bc})\rangle$ for each $\tilde\Theta_{ab}$ ($x$-axis) and $\tilde\Theta_{bc}$ ($y$-axis) (optimized via local unitaries). $Q_{SEP}$  is always greater zero, indeed even $\geq\frac{1}{2}$, thus proving entanglement for all possible decay scenarios. The quantities detecting genuine multipartite entanglement $Q_{W},Q_{GHZ}$ are greater than zero, thus detecting genuine multipartite entanglement, however, their values differ.}
	\label{sepforpure}
\end{figure*}

Considering the polarisation of three photons the state under investigation is a tripartite qubit system that we discuss now with respect to its entanglement features of the polarisation degrees independent of the kinematic constraints.  Any entanglement of a tripartite state can be classified according to the $k$-separability~\cite{Horodecki} (for a more recent overview over the subtleties concerning the classification of multipartite states see e.g. Ref.~\cite{Chruscinski}). If a pure $n$-partite state can be written in the form
\beq
|\psi\rangle=|\phi_1\rangle\otimes |\phi_2\rangle\otimes\dots\otimes|\phi_k\rangle
\eeq
with $k\leq n$, it is called for $k=n$ \textit{fully separable}, for $1<k<n$ \textit{partially separable} (\textit{$k$-separable}) or for $k=1$ \textit{fully entangled} (\textit{$1$-separable}). There is a straight forward extension to mixed states, i.e.
if a mixed state can be written as a convex combination of at least $k$-separable states, i.e. ($p_i\geq 0$)
\beq
\rho=\sum_i p_i\; \rho_i^1\otimes\rho_i^2\otimes\dots\otimes \rho_i^k
\eeq
with $k\leq n$, it is classified as in the case for pure states. Note that $1$-separable states are also called \textit{genuinely multipartite entangled} states and these states are the most interesting ones with respect to outperforming algorithms exploiting classical physics.

This classification is certainly not fine enough. Already for the simplest case, three qubits, we have two physically very different subclasses of $1$-separable states or genuinely multipartite entangled states:
The GHZ-states (GHZ\dots Greenberger, Horne, Zeilinger)
\beq |GHZ\rangle=\frac{1}{\sqrt{2}}\left(|000\rangle+|111\rangle\right)\eeq
and the Dicke states ($W$ states called in the case of three qubits), e.g.,
\beq |W\rangle=\frac{1}{\sqrt{3}}\left(|001\rangle+|010\rangle+|100\rangle\right)\;.\eeq
Both are obviously a particular generalization of the maximally entangled  Bell states, but there physical properties are very different. For example, if one subsystem is traced out the entanglement is fully lost in the case of $GHZ$-type of entanglement, in contrast to the $W$-type of entanglement, where the subsystems are entangled. It has been shown that the $GHZ$-type of entanglement can be utilized for multipartite quantum cryptography~\cite{SHH,Karlsson,HBB} whereas for Dicke-type entanglement no such schemes have been found that outperforms bipartite entangled systems. Dicke-type of entanglement is often present in condensed matter systems~\cite{Dicke} or are produced by a double down conversion process resulting in four genuinely multipartite entangled photons~\cite{meoam,zeilingeroam}. For single neutrons in an interferometric setup three degrees of freedom can be engineered, i.e. spin, path and energy, for which both types of genuine multipartite entanglement have been generated experimentally~\cite{meneutron}. Recently, also atoms in a solid have been proven to be genuinely multipartite entangled~\cite{gmesolid1,gmesolid2}. Topological and nematic phase transitions in spin chains are shown to be ruled by genuine multipartite entanglement~\cite{phasetransition1,phasetransition2,phasetransition3}, even if bipartite entanglement dies out~\cite{phasetransition4}.

Our first aim is to analyze the entanglement of the three photons resulting from the ortho-positronium decay.
Without loss of generality we can set $\Phi_{plane}=0$ and $s_{\hat{\vec{n}}}=0$ since the entanglement properties do not depend on local unitaries (if not mixed). Thus the state under investigation depends only on the two azimuth angles
\beq |\psi_{pure}(\tilde\Theta_{ab},\tilde\Theta_{bc})\rangle&=&\hat{\mathcal{R}}_{pol}(\tilde\Theta_{ab},\tilde\Theta_{bc}) \;|\Psi\rangle_{abc}\;.
\eeq
For pure states well-established bipartite entanglement criteria serve the purpose to reveal all the entanglement. In our case, however, we will stick to the HMGH-framework~\cite{HMGH}, which we apply later to mixed states (discussed in the section~\ref{mixedness}). This framework connects local observables, density matrix elements, to distinct types of entanglement.

In Ref.~\cite{HMGH} it was proven that the following criterion holds for all $k$-separable states $\rho$
\beq\label{k-separabilitycriterion} 
\lefteqn{Q_k(\rho)\;=}\\
&&\sqrt{\bra{\chi}\rho^{\otimes 2} P_{total}\ket{\chi}} - \sum_{\{\alpha\}}\left(\prod_{i=1}^k \bra{\chi}P_{\alpha_i}^\dagger \rho^{\otimes 2} P_{\alpha_i} \ket{\chi}\right)^{\frac{1}{2k}} \leq 0\;,\nonumber\eeq
where $\ket{\chi}=\ket{\chi_1}\otimes\ket{\chi_2}$ is an arbitrary fully separable state, $P_{\alpha_i}$ is a permutation operator permuting the $\alpha_i$-th elements of $\ket{\chi_1}$ and $\ket{\chi_2}$ and the sum runs over all $k$-partitions $\{\alpha\}$. And the total permutation  acts as
$ P_{total}\ket{\chi_1}\otimes\ket{\chi_2} = \ket{\chi_2}\otimes\ket{\chi_1}$. Obviously, if this inequality is violated the state $\rho$ cannot be $k$-separable. Note that the reverse argument does not hold since a non-violation does not necessarily imply $k$-separability. Consequently, these criteria are necessary but not sufficient criteria for $k$-separability. Since the above criteria obviously depend on the choices of the fully separably $\ket{\chi}$ and the chosen basis of $\rho$ one always has to optimize over local unitaries in order to obtain the optimum.

For three qubits a value of $Q_{k=3}(\rho)$ greater zero detects entanglement of $\rho$ and a value of $Q_{k=2}(\rho)$ greater zero detects the state to be genuinely multipartite entangled. It turns also out that $Q_{k=2}(\rho)$ is the one that gives the highest value for the $GHZ$-state, i.e. $Q_{k=2}(|GHZ\rangle)=1$. In strong contrast to the $W$ states which give $Q_{k=2}(|W\rangle)=0.629$. In the following we denote therefore this criterion by $Q_{GHZ}:=Q_{k=2}$. And by $Q_{SEP}:=Q_{k=3}$ the criterion detecting entanglement but not necessarily genuine multipartite entanglement. Explicitly, we can rewrite the criteria which detects entanglement if the value is greater than zero by
\begin{widetext}
\beq\label{bipartitecriterion}
Q_{SEP}(\rho)&=& 2\cdot|\langle 000|\rho|111\rangle|-2\left(\langle 001|\rho|001\rangle\langle 010|\rho|010\rangle\langle 011|\rho|011\rangle\langle 100|\rho|100\rangle\langle 101|\rho|101\rangle\langle 110|\rho|110\rangle\right)^{\frac{1}{6}}\;.
\eeq
The criterion that detects genuine multipartite entanglement if the value is greater than zero and maximizes for any $GHZ$-state rewrites to
\beq\label{criteriondetailed}
Q_{GHZ}(\rho)&=&2\cdot\left(|\langle 000|\rho|111\rangle|-\sqrt{\langle 110|\rho|110\rangle\langle 001|\rho|001\rangle}-\sqrt{\langle 101|\rho|101 \rangle\langle 010|\rho|010\rangle}-\sqrt{\langle 011|\rho|011 \rangle\langle 100|\rho|100\rangle}\right)\;.\nonumber\\
\eeq
\end{widetext}
This formulation reveals the very working of the criteria, i.e. that the only off-diagonal element of these criteria are exactly the only non-zero off-diagonal element of the $GHZ$-state and the negative terms are diagonal elements of $\rho$ which are all zero in the case of the $GHZ$-state.

A criterion to optimize for the $W$-type of entanglement can be also derived via the HMGH-framework~\cite{HMGH}. The same strategy as above can be used, i.e. choosing the non-zero elements of the $W$-state, i.e. $|001\rangle,|010\rangle$ or $|100\rangle$ for the fully separable state $\ket{\chi_1}$ and $\ket{\chi_2}$. Since we have now three combinations we can add these three inequalities to have a symmetric criterion. In Ref.~\cite{Dickeme}, however, it was shown that one can obtain a stricter inequality if one adds a further constraint coming from the positivity condition. For $3$-particles with two degrees of freedoms the following criterion detects genuine multipartite entanglement if greater zero and attains its maximal value for the $W$-state
\begin{widetext}
\beq\label{Dicketripartite}
Q_{W}(\rho)&=& 2 |\langle 001|\rho|010\rangle|+2 |\langle 001|\rho|100\rangle|+2|\langle 010|\rho|100\rangle|-\biggl(\langle 001|\rho|001\rangle+\langle 010|\rho|010\rangle+\langle 100|\rho|100\rangle\nonumber\\
 &&+ 2 \sqrt{\langle 000|\rho|000\rangle\cdot \langle 011|\rho|011\rangle}+2 \sqrt{\langle 000|\rho|000\rangle\cdot \langle 101|\rho|101\rangle}+2 \sqrt{\langle 000|\rho|000\rangle\cdot \langle 110|\rho|110\rangle}\biggr)\;.
\eeq
\end{widetext}
The positive terms are the only non-zero off-diagonal terms of the $W$-state whereas the negative terms are only diagonal terms that are zero in the case of the $W$-state. Therefore, this criterion gives the maximum value for the $W$-state. In strong contrast to $GHZ$-states which obtain the value $Q_{W}(|GHZ\rangle)=\frac{3}{4}$. The separability criterion for both genuinely multipartite entangled states is $Q_{SEP}(|GHZ\rangle)=1$ and $Q_{SEP}(|W\rangle)=0.62$. A summary --including the positronium case-- can be found in TABLE~\ref{table}.

\begin{center}
\begin{table*}
\begin{tabular}{c|ccc}
3 Qubits&$Q_{GHZ}$& $Q_{W}$& $Q_{SEP}$ \\
\hline
$|GHZ\rangle$&$1$&$\frac{3}{4}$&$1$\\
$|W\rangle$&$0.628$&$1$&$\frac{2}{3}$\\
$\max_{{\tiny\tilde\Theta_{ab},\tilde\Theta_{bc}}}|\psi_{pure}\rangle$&$0.76$ &$0.83$&$0.89$\\
&{\tiny$(\tilde\Theta_{ab}=\frac{15 \pi}{8},\tilde\Theta_{bc}=\frac{\pi}{4})$}&\tiny{$ (\tilde\Theta_{ab}=\frac{15 \pi}{8},\tilde\Theta_{bc}=\frac{\pi}{4})$}&{\tiny$(\tilde\Theta_{ab}=\frac{\pi}{16},\tilde\Theta_{bc}=\frac{\pi}{16})$}\\
$|\psi_{pure}(\frac{2\pi}{3},\frac{2\pi}{3})\rangle$& $0.58$&$0.67$&$0.67$\\
$\rho_{mixed}(\frac{1}{3},0)(\frac{2\pi}{3},\frac{2\pi}{3})$&$0$&$0.5$&$0.17$\\
\end{tabular}\caption{The optimized values of the three entanglement criteria for different three qubit states.}\label{table}
\end{table*}
\end{center}

Obviously these criteria are measurable by local observables since they depend only on density matrix elements, for example
\beq
\langle 000|\rho|111\rangle&=&\langle \sigma_x\otimes\sigma_x\otimes\sigma_x\rangle_{\rho}-\langle \sigma_x\otimes\sigma_y\otimes\sigma_y\rangle_{\rho}\nonumber\\
&&-\langle \sigma_y\otimes\sigma_x\otimes\sigma_y\rangle_{\rho}-\langle \sigma_y\otimes\sigma_y\otimes\sigma_x\rangle_{\rho}\nonumber\\
&&-i\left(\langle \sigma_x\otimes\sigma_x\otimes\sigma_y\rangle_{\rho}+\langle \sigma_x\otimes\sigma_y\otimes\sigma_x\rangle_{\rho}\right.\nonumber\\
&&\left.+\langle \sigma_y\otimes\sigma_x\otimes\sigma_x\rangle_{\rho}-\langle \sigma_y\otimes\sigma_y\otimes\sigma_y\rangle_{\rho}\right)\;,\nonumber\\
\eeq
where $\sigma_i$ are the Pauli matrices. This makes the criteria very experimenter friendly since they are attainable by local measurements only and do not need state tomography.

Deriving $Q_{SEP}$ for the three-photon pure state we find $Q_{SEP}(|\psi_{pure}(\tilde\Theta_{ab},\tilde\Theta_{bc})\rangle)\geq\frac{1}{2}$, i.e. a positive value for all possible angles. Thus proving that the three-photon pure states resulting from the decay of positronium are always entangled. The maximum equals to $0.89$ and is reached for $\tilde\Theta_{ab}=\frac{\pi}{16}, \tilde\Theta_{bc}=\frac{\pi}{16}$ in the non-physical area. The contour plot in Fig.~\ref{sepforpure}~(a) shows the details. Whereas the three-photon state is entangled for all possible decay configurations $\{\tilde\Theta_{ab},\tilde\Theta_{bc}\}$, this holds surprisingly also true for genuine multipartite entanglement detected by $Q_{GHZ}$ or by $Q_{W}$. Both criteria differ only in the amount of the violation of the inequality, see  Fig.~\ref{sepforpure}~(b) and (c).

Summarizing, without the kinematic constraints on the polarisation degrees of freedom, $R_{pol}(\tilde\Theta_{ab},\tilde\Theta_{bc})$,  the state in the decay process of the positronium would be a pure $GHZ$-state. These constrains result in a dependence of entanglement on the decay angles obeying the indistinguishability of the individual photons.

The next section in which we analyse the distribution of the bipartite entanglement between the individual photons will answer how these kinematic constraints onto the polarisation degrees of freedom prefers different types of genuine multipartite entanglement.

\section{Entanglement of the reduced system}\label{reduced}

\begin{figure}
\centering{\includegraphics[width=0.3\textwidth]{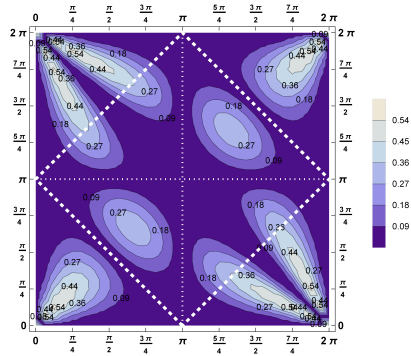}}
\caption{(Color online) This contour plot shows the tangle minus the two concurrences, $\tau_{i|jk}-C(\rho_{ij})^2-C(\rho_{ij})^2$, which is equal for any permutation of the three photons.}
	\label{tangle}
\end{figure}

Now we want to investigate how the entanglement is distributed among the individual photons $a,b$ and $c$. From equation~(\ref{reinerzustand}) it is obvious that without the operator $\hat{\mathcal{R}}_{pol}(\tilde\Theta_{ab},\tilde\Theta_{bc})$ the reduced state is an equal mixture of two Bell states which is a separable state. The kinematic operator weights not the two Bell states separately but each contribution individually. By that the invariance under permutation of the photons is lost, i.e. the three reduced states $\rho_{ab},\rho_{ac},\rho_{bc}$ differ as well as its entanglement content measured by concurrence differs. Concurrence is an analytically computable entanglement measure for qubit-qubit entanglement. For pure states it simplifies to $C(|\psi\rangle)=\sqrt{2(1-Tr(\rho_i))}$ where $\rho_i$ is the partial trace of $|\psi\rangle$ with respect to the subsystem $i$. For a mixed state concurrence is defined by the convex roof, i.e.
\beq
\lefteqn{C(\rho)\;=}\\
&&\min_{p_i, |\psi_i\rangle}\left\{\sum p_i C(\psi_i) | \sum p_i |\psi_i\rangle\langle \psi_i|=\rho \;\textrm{with}\; p_i\geq 0\right\}\;.\nonumber
\eeq
For bipartite qubits it has been shown that the convex sum equals the value obtained by computing the eigenvalues of $\sqrt{Tr(\rho\sigma_2\otimes\sigma_2\rho^*\sigma_2\otimes\sigma_2)}$ and taking the maximal eigenvalue minus the remaining ones.

Since our total three-qubit system is a pure state we can directly answer the question how much bipartite entanglement photon $a$ shares with photon $b$ and photon $c$. Obviously, if photons $b$ and $c$ are in a maximal entangled state they have to be separable to photon $a$. Therefor, the entanglement $a$ can share with $bc$ limits the entanglement of $bc$.
This can be quantified with the help of famous Coffman-Kondu-Wootters tangle $\tau_{a|bc}$~\cite{CKW}, i.e.
\beq
C(\rho_{ab})^2+C(\rho_{ac})^2\leq\;\tau_{a|bc}(\psi_{abc})\;:=\; 4 \det\rho_{a}\;.
\eeq
For the $GHZ$-state the reduced states are separable states thus the concurrence is zero, whereas the tangle is maximal. The difference between the right hand side and the left hand side is maximal. Whereas in the case of the $W$-state the reduced states have a value of $C=\frac{2}{3}$ and the tangle equals $\tau=\frac{8}{9}$, i.e. the difference is zero. Thus entanglement is distributed also among the subsystems in the case of a $W$-state, whereas for $GHZ$-states no entanglement can be found in the subsystems. Thus the tangle minus the two concurrences quantifies the difference between genuine multipartite entanglement of the $GHZ$-type of entanglement and the $W$-type of entanglement.

Let us note here, however, another subtle point of multipartite entanglement. Obviously, if a GHZ-state as given in equation~(\ref{reinerzustand}) is considered, a measurement of one photon in the circular polarized basis ($\{|+\rangle,|-\rangle\}$) leads to a separable state for the two remaining photons, i.e. $|++\rangle$ or $|--\rangle$. If the photon is instead measured in the linear polarized basis ($\{|0\rangle,|1\rangle\}$) the remaining two photons are for the result ``0'' in the Bell state $|\phi^-\rangle$ or for the outcome ``1'' in the Bell state $|\psi^+\rangle$, i.e. clearly maximally entangled. This perfect correlation between the polarization state of one photon and the entangled state of the two photons implies, under the Einstein-Podolsky-Rosen premises of realism and no action at a distance, that the entangled state of the two photons must represent an element of reality. Whereas the individual photons, which have no well-defined properties, do not correspond to such elements. For a realist this is a surprising feature. In the first scenario the two photons contain individually an element of reality, which is more satisfactory for a realist. Thus by the specific kind of measurement, projecting on linearly or circularly polarized photons, the properties of the two photons and their reality content is switched between entanglement and separability. This can also be understood from the fact that a particular factorisation per se is not favoured over another one, no partition has ontologically a superior status over any other one, there is perfect democracy. However, a measurement or a physical process makes a choice.

Thus let us come back to the $3$-photon decay resulting from ortho-positronium. Though the individual concurrences for a given setup $(\tilde\Theta_{ab},\tilde\Theta_{bc})$ differ, the difference of the tangle minus the two squared concurrences, $\tau_{i|jk}-C(\rho_{ij})^2-C(\rho_{ij})^2$, has to be equal for all possible permutations of the three photons. This expresses that the kinematics chosen by the decay process  chooses the respected bipartite entanglement to be larger or smaller compared to the other bipartitions, however, the total amount entanglement does not depend on the individual choice. The largest difference between the individual bipartite entanglement and the one shared with both remaining photons overlaps with the regions for which genuine multipartite entanglement maximizes, showing that the $W$-type of entanglement is more resistent against the specific setup (angles) chosen by the decay process respecting the indistinguishability of the photons. The details are plotted in Fig.~\ref{tangle}.

Note that a decaying system can be viewed as an open quantum process~\cite{HiesmayrOpen}, i.e. an interaction with an environment plays the role of the choice of measurement of the system of interest.

\section{Entanglement properties of the mixed  $3$-photon-states}\label{mixedness}

\begin{figure*}
\centering{(a)\includegraphics[width=0.3\textwidth]{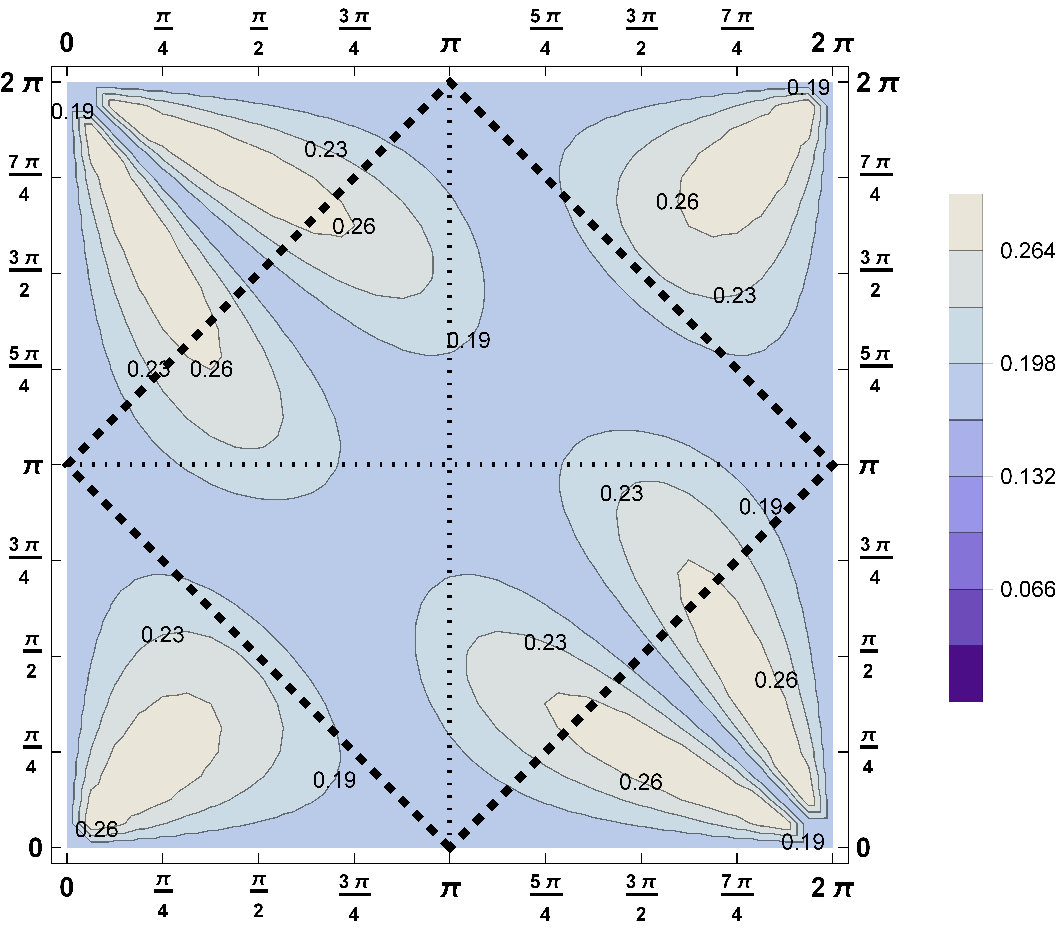}(b)\includegraphics[width=0.3\textwidth]{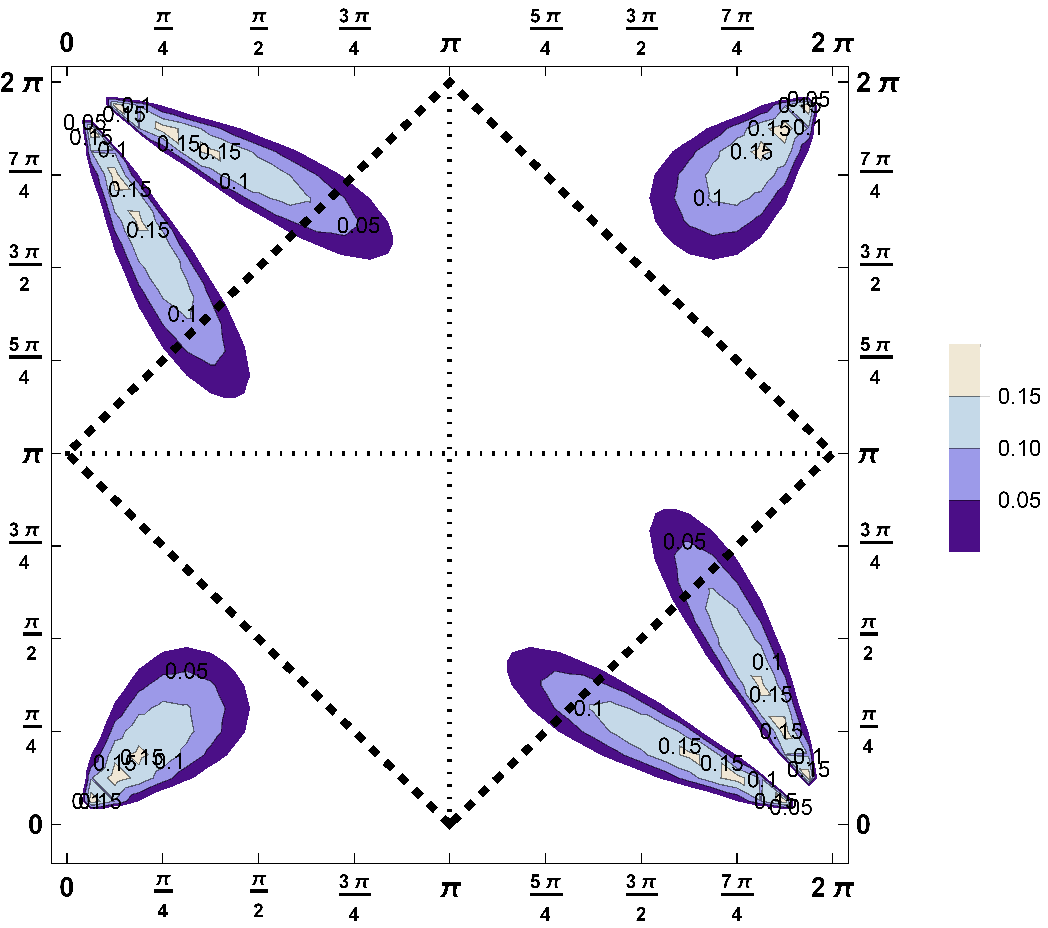}
(c)\includegraphics[width=0.3\textwidth]{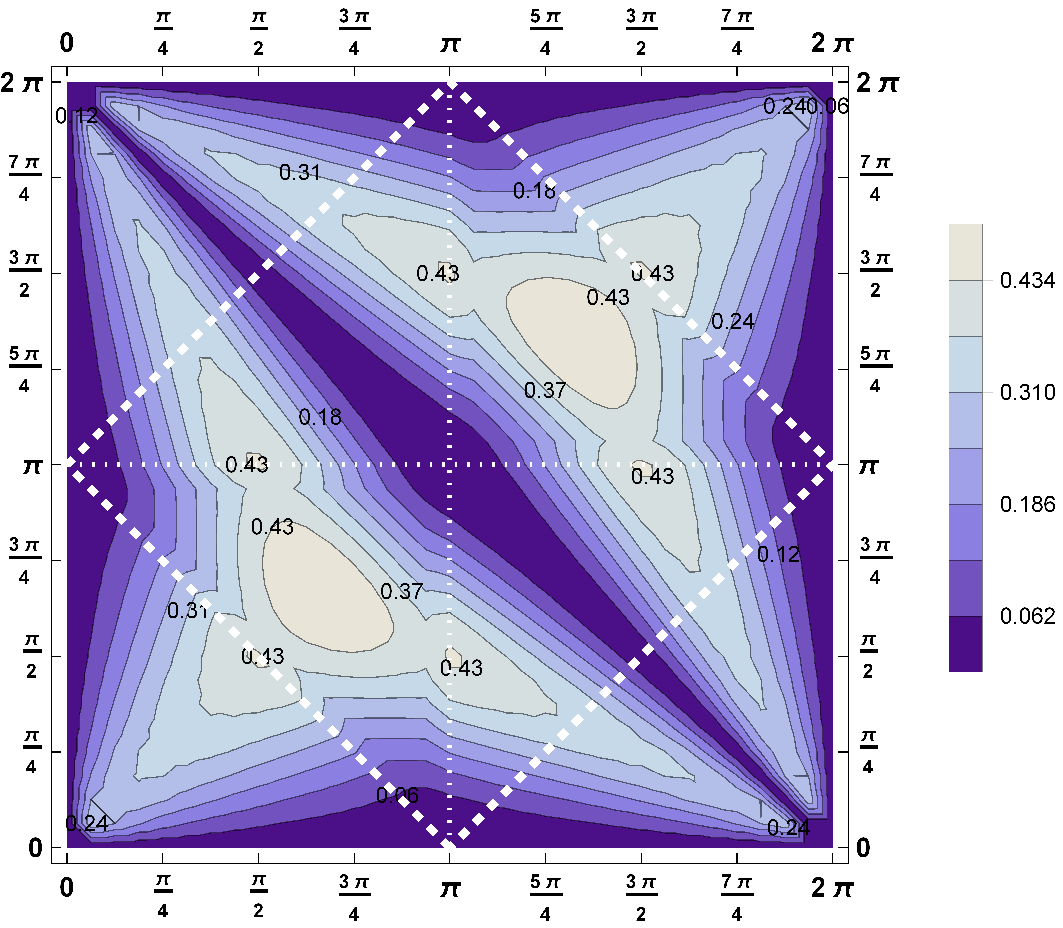}}
\caption{(Color online) These three contour plots show (a) $Q_{SEP}$, (b)  $Q_{GHZ}$ and (c) $Q_{W}$ for the state mixed equally between all three possible quantum states $s_{\hat{\vec{n}}}=0,+1,-1$, equation~\ref{mixedstate}. Still genuine multipartite entanglement is revealed for some scenarios $(\tilde\Theta_{ab},\tilde\Theta_{bc})$. The criterion $Q_W$ detecting $W$-type of genuine multipartite entanglement is by far more sensitive to reveal genuine multipartite entanglement.}
	\label{figuremixed}
\end{figure*}

In case spin is not a proper quantum number all three possible spin eigenstates $s_{\hat{\vec{n}}}=0,+1,-1$ are equally probable, then the resulting state is
\beq\label{mixedstate}
\rho_{mixed}(p,\Phi_{plane})&=& p\; |\Psi_{s_{\hat{\vec{n}}}=0}\rangle\langle\Psi_{s_{\hat{\vec{n}}}=0}|\nonumber\\
&&+\frac{1-p}{2}\; |\Psi_{s_{\hat{\vec{n}}}=+ 1}\rangle\langle\Psi_{s_{\hat{\vec{n}}}=+ 1}|\nonumber\\
&&+\frac{1-p}{2}\; |\Psi_{s_{\hat{\vec{n}}}=-1}\rangle\langle\Psi_{s_{\hat{\vec{n}}}=- 1}|
\eeq
with $p=\frac{1}{3}$. Computing the mixedness $Tr\rho^2$ we find for $p=1,\frac{1}{3}$ no dependence on $\Phi_{plane}$. This is also the case for the entanglement properties of the state. The three criteria $Q_{SEP},Q_{GHZ}$ and $Q_{W}$ are presented in Fig.~\ref{figuremixed}. Remarkably, entanglement is not lost for any setup, however, genuine multipartite entanglement is more sensitive to the angles. Concerning the physical attainable region we find that $Q_{SEP}$ attains in good approximation a constant value ($Q_{SEP}=[0.17,0.2]$) which shows a kind of symmetrization in the sense that the difference between the relevant off diagonal element and the sum of the relevant diagonal elements of the density matrix is constant, see equation~(\ref{bipartitecriterion}), however, it is strictly non-positive for $Q_{GHZ}$. In strong contrast, $Q_W$ reveals differences in the entanglement properties since it still varies strongly with the angles. Consequently, this criterion reveals refined properties of the system under investigation. This proves that entanglement properties of the state can be revealed in the highly mixed scenario and gives the hope that biological properties of the system may be revealable.

In summary, $GHZ$-type of entanglement can no longer be found in a physically available region, however, $W$-type of entanglement is robust against this mixing. Particularly, in the fully symmetric case
$\tilde\Theta_{ab}=\tilde\Theta_{bc}=\frac{2 \pi}{3}$ we find a local maximum with the value $Q_W=0.5$. This shows that the dynamics of the decay process does not wash out fully the entanglement features and favours Dicke-type of entanglement over GHZ-type of entanglement. Thus the decay process favours a symmetrization among the three photons enabling bipartite entanglement.

\section{Summary and Outlook}

Monitoring metabolic processes as well as distinct physical reactions related to chemical processes are key ingredients to explore nature and its very working. We analyze the entanglement properties in the polarisation degrees of freedom of three photons resulting from the decay of ortho--positronium in the full parameter space. For that we use the HMGH-framework which allows to detect entanglement in addition to refinements such as $W$-type or more generally $Dicke$-type of entanglement versus $GHZ$-type of entanglement. The framework provides non-linear entanglement witnesses based on local observables, i.e. does not need full information on the state that is in many cases not attainable.

In particular, for a definite spin value of ortho-positronium we find that entanglement and even stronger genuine multipartite entanglement is present for the full parameter space. Surprisingly, in the mixed scenario entanglement as well as genuine multipartite entanglement are not lost, however, only $W$-type of genuine multipartite entanglement is detectable. This is due to the interplay of the kinematics of the three-body particle decay and the Bose-symmetry constraining the entanglement properties to favor $W$-type over $GHZ$-type of genuine multipartite entanglement. Furthermore, whereas the criterion detecting entanglement, $Q_{SEP}$, is in good approximation constant over the physical relevant region, this is not the case for the specific criterion $Q_{W}$, it reveals different properties in dependence of the decay angles due to a specific ``\textit{symmetrisation process}'' as an effect of the decay process. Since the mixing does not destroy entanglement per se and genuine multipartite entanglement is shown to be still dependent on the angles, this  proves that entanglement can be related to physical processes and gives the hope that entanglement will maybe be related to real biological processes.

J-PET, a Positron-Emission-Tomograph, relies on new technology enabling three-photon tomography~\cite{PawelPatent,Dariajpet,JPETGajus}. This is due to a new detector scheme based on plastic scintillators~\cite{JPET1}, novel digital sampling electronics~\cite{JPET2,JPET3} and a development of trilateration-based reconstruction~\cite{JPETGajus}. Consequently, J-PET gives a possibility to determine the linear polarization of high energy photons via the registration of the direction of the photon before and after its Compton scattering~\cite{JPET3}.
It allows measuring the correlations of photons with superior time and angular resolutions via Compton scattering. It is out of the scope of this contribution to compute how the entanglement of the three photons can be revealed by the three Compton scattered photons, we will tackle this problem in a future work.

\vspace{2cm}
\textbf{Acknowledgement:} B.C.H. acknowledges gratefully the Austrian
Science Fund (FWF-P26783) and P.M. acknowledges support by the National Science Centre through the grant No. 2016/21/B/ST2/01222.


\begin{thebibliography}{9}


\bibitem{PETdevelopment}
P. Slomka, T. Pan and G. Germano, \textit{Recent Advances and Future Progress in PET Instrumentation}, Semin. Nucl. Med. 46, 5 (2016).

\bibitem{PETdevelopment2}
I. Rausch, H. H. Quick, J. Cal-Gonzalez, B. Sattler, R. Boellaard and T. Beyer, \textit{Technical and instrumentational foundations of PET/MRI}, European Journal of Radiology. In Press. Available at: http://dx.doi.org/10.1016/j.ejrad.2017.04.004.

\bibitem{PETdevelopment3}
S. Vandenberghe, P. K. Marsden, \textit{PET-MRI: a review of challanges and solutions in the development of integrated multimodality imaging},
Phys. Med. Biol. 60, R115 (2015).


\bibitem{PETdevelopment4}
M. Conti, \textit{Focus on time-of-flight PET: the benefits of improved time resolution}, Eur. J. Nucl. Med. Mol. Imaging 38, 1147 (2011).


\bibitem{PETdevelopment5}
J. S. Karp et al., \textit{Benefit of Time-of-Flight in PET: Experimental and Clinical Results}, J. Nucl. Med. 49, 462 (2008).


\bibitem{PETdevelopment6}
J. L. Humm, A. Rosenfeld,  A. Del Guerra, \textit{From PET detectors to PET scanners}, Eur. J. Nucl. Med. Mol. Imaging 30, 1574 (2003).


\bibitem{PETdevelopment7}
W. W. Moses and S. E. Derenzo, \textit{Prospects for Time-of-Flight PET using LSO scintillator}, IEEE Trans. Nucl. Sci. 46, 474 (1999).


\bibitem{positronium1}
M. Deutsch, \textit{ Three-Quantum Decay of Positronium}, Phys. Rev. 82, 455 (1951).

\bibitem{positronium2}
A. H. Al-Ramadhan, D. W. Gidley, \textit{New precision measurement of the decay rate of singlet positronium}, Phys. Rev. Lett. 72, 1632 (1994).

\bibitem{positronium3}
R. S. Vallery, P. W. Zitzewitz, D. W. Gidley, \textit{Resolution of the Orthopositronium-Lifetime Puzzle}, Phys. Rev. Lett. 90, 203402 (2003).


\bibitem{positronium4}
D. B. Cassidy, P. Crivelli, T. H. Hisakado, L. Liszkay, V. E. Meligne, P. Perez, H. W.K. Tom, A. P. Mills Jr,
\textit{Positronium cooling in porous silica measured via Doppler spectroscopy}, Phys.Rev. A 81, 012715 (2010).



\bibitem{PET3gammeprobability}
S. V. Stepanov, D. S. Zvezhinskiy, G. Duplatre, V. M. Byakov, Y. Y. Batskikh and P. S. Stepanov, \textit{Incorporation of the Magnetic Quenching Effect into the Blob Model of Ps Formation. Finite Sized Ps in a Potential Well}, Materials Science Forum, Vol. 666, 109 (2011).

\bibitem{tissueprobability}
M.D. Harpen, \textit{Positronium: review of symmetry, conserved quantities and decay for the radiological physicist}, Med. Phys. 31, 57 (2004).



\bibitem{JPET1}
P. Moskal et al., \textit{Time resolution of the plastic scintillator strips with matrix photomultiplier readout for J-PET tomography}, Phys. Med. Biol. 61, 2025 (2016).

\bibitem{JPET2}
P. Moskal et al., \textit{A novel method for the line-of-response and time-of-flight reconstruction in TOF-PET detectors based on a library of synchronized model signals}, Nucl. Instr. Meth. A 775, 54 (2015).

\bibitem{JPET3} P. Moskal and the J-PET collaboration, \textit{Potential of the J-PET detector for studies of discrete symmetries in decays of positronium atom - a purely leptonic system}, Acta Phys. Polon. B 47, 509 (2016).


\bibitem{PawelPatent}
P. Moskal et al., \textit{TOF-PET tomograph and a method of imaging using a TOF-PET tomograph, based on a probability of production and lifetime of a positronium}, Patent Application: PCT/EP2014/068374\;.


\bibitem{Dariajpet}
D. Kaminska et al., \textit{A feasibility study of ortho-positronium decays measurement
with the J-PET scanner based on plastic scintillators}, Eur. Phys. J. C 76, 445 (2016).

\bibitem{JPETGajus}
A. Gajos et al., \textit{Trilateration-based reconstruction of ortho-positronium decays into three photons with the J-PET detector}, Nucl. Instr. and Meth. A 819, 54 (2016).



\bibitem{Giacomo}
G. Guarnieri, M. Motta and L. Lanz, \textit{Single-photon observables and preparation uncertainty relations}, J.Phys. A: Math. Theor. 46, 265302 (2015).


\bibitem{bio1}
M. Sarovar, A. Ishizaki, G. R. Fleming and K. B. Whaley, \textit{Quantum entanglement in photosynthetic light-harvesting complexes}, Nature Physics, 6, 462 (2010).

\bibitem{bio2}
J. Cai, G. G. Guerreschi and H. J. Briegel, \textit{Quantum Control and Entanglement in a Chemical Compass}, Phys. Rev. Lett 104, 220502 (2010).

\bibitem{bio3} E. M. Gauger, E. Rieper, J.J.L. Morton, S.C. Benjamin and V. Vedral, \textit{Sustained Quantum Coherence and Entanglement in the Avian Compass}, Phys. Rev. Lett. 106, 040503 (2011).

\bibitem{bio4}
M.I. Franco, L. Turin, A. Mershin and E.M.C. Skoulakis, \textit{Molecular vibration-sensing component in Drosophila melanogaster olfaction}, Proc. Natl. Acad. Sci. Am. 108, 3797 (2011).

\bibitem{JPETEntanglementGeneration}
B.C. Hiesmayr and J-PET collaboration, \textit{Entanglement generation in the decay of positronium}, in preparation.

\bibitem{Acin}
A. Acin, J. I. Latorre, and P. Pascual, \textit{Three-party entanglement from positronium}, Phys. Rev. A 63, 042107 (2001).

\bibitem{Horodecki}
M. Horodecki, P. Horodecki and R. Horodecki, \textit{Separability of Mixed States: Necessary and Sufficient Conditions}, Phys. Lett. A 223, 1 (1996).

\bibitem{Chruscinski}
D. Chruściński and G. Sarbicki, \textit{Entanglement witnesses: construction, analysis and classification}, J. Phys. A: Math. Theor. 47, 483001 (2014).



\bibitem{SHH}
St. Schauer, M. Huber and B. C. Hiesmayr, \textit{Experimentally Feasible Security Check for n-qubit Quantum Secret Sharing}, Phys. Rev. A 82, 062311 (2010).

\bibitem{Karlsson}
A. Karlsson, M. Koashi and N. Imoto, \textit{Quantum Entanglement for Secret Sharing and Secret Splitting}, Phys. Rev. A 59, 162 (1999).

\bibitem{HBB}
M. Hillery, V. Buzek, and A. Berthiaume, \textit{Quantum Secret Sharing}, Phys. Rev. A
59, 1829 (1999).

\bibitem{Dicke}
R. H. Dicke, \textit{Coherence in Spontaneous Radiation Processes}, Phys. Rev. 93, 99 (1954).

\bibitem{meoam}
B. C. Hiesmayr, M. J. A. de Dood and W. L\"offler, \textit{Four-photon orbital angular momentum entanglement},
Phys. Rev. Lett. 116, 073601 (2016).

\bibitem{zeilingeroam}
M. Krenn, M. Malik, M. Erhard and A. Zeilinger, \textit{Orbital angular momentum of photons and the entanglement of Laguerre-Gaussian modes},
Philosophical Transactions of the Royal Society A 375, 20150442 (2016).


\bibitem{meneutron}
D. Erd\"osi, M. Huber, B. C. Hiesmayr and Y. Hasegawa, \textit{Proving the Generation of Genuine Multipartite Entanglement in a Single-Neutron Interferometer Experiment}, New J. Phys. 15, 023033 (2013).

\bibitem{gmesolid1}
F. Fr\"owis, P. C. Strassmann, A. Tiranov, C. Gut, J. Lavoie, N. Brunner, F. Bussieres, M. Afzelius and N. Gisin, \textit{Experimental certification of millions of genuinely entangled atoms in a solid},  arXiv:1703.04704.

\bibitem{gmesolid2}
P. Zarkeshian, C. Deshmukh, N. Sinclair, S.K. Goyal, G.H. Aguilar, P. Lefebvre, M. Grimau Puigibert, V.B. Verma, F. Marsili, M.D. Shaw, S.W. Nam, K. Heshami, D. Oblak, W. Tittel and C. Simon, \textit{Entanglement between more than two hundred macroscopic atomic ensembles in a solid}, arXiv:1703.04709.

\bibitem{phasetransition1}
S. M. Giampaolo and B.C. Hiesmayr, \textit{Topological and nematic ordered phases in many-body cluster-Ising models}, Phys. Rev. A 92, 012306 (2015).

\bibitem{phasetransition2}
S. M. Giampaolo and G. Zonzo, \textit{Quench of a symmetry-broken ground state}, Phys. Rev. A, 95, 012121 (2017).

\bibitem{phasetransition3}
T. E. Lee, J.N. Joglekar, and P. Richerme, \textit{String order via Floquet interactions in atomic systems},
Phys. Rev. A 94, 023610 (2017).

\bibitem{phasetransition4}
Y.-Y. Zhang, X.-Y. Chen, S. He and Q.-H. Chen, \textit{Analytical solutions and genuine multipartite entanglement of the three-qubit Dicke model}, Phys. Rev. A 94, 012317 (2016).


\bibitem{HMGH}
M. Huber, F. Mintert, A. Gabriel and B.C. Hiesmayr, \textit{Detection of high-dimensional genuine multi-partite entanglement of mixed states}, Phys. Rev. Lett. 104, 210501 (2010).


\bibitem{Dickeme}
M. Huber, P. Erker, H. Schimpf, A. Gabriel and B.C. Hiesmayr, \textit{Experimentally feasible set of criteria detecting genuine multipartite entanglement in n-qubit Dicke states and in higher dimensional systems}, Phys. Rev. A 83, 040301(R) (2011); Phys. Rev. A 84, 039906(E) (2011).

\bibitem{CKW}
V. Coffman, J. Kundo, W.K. Wootters, \textit{Distributed entanglement}, Phys. Rev. A 61, 052306 (2000).

\bibitem{HiesmayrOpen}
R. A. Bertlmann, W. Grimus and B.C. Hiesmayr, \textit{An open--quantum--system formulation of particle decay}, Phys. Rev. A 73, 054101 (2006).


\end{thebibliography}
\end{document}